\documentclass[sn-nature]{sn-jnl}
\usepackage{lineno}

\usepackage{graphicx}%
\usepackage{multirow}%
\usepackage{amsmath,amssymb,amsfonts}%
\usepackage{amsthm}%
\usepackage{mathrsfs}%
\usepackage[title]{appendix}%
\usepackage{xcolor}%
\usepackage{textcomp}%
\usepackage{manyfoot}%
\usepackage{booktabs}%
\usepackage{algorithm}%
\usepackage{algorithmicx}%
\usepackage{algpseudocode}%
\usepackage{listings}%

\usepackage{dcolumn}

\usepackage{bm}
\usepackage[version=4]{mhchem}
\usepackage{xr-hyper}
\usepackage{siunitx}

\usepackage{myxrhyper}
\myexternaldocument[sm-]{build}{sm}

\newcommand*{\abs}[1]{\left\lvert#1\right\rvert}
\newcommand*{\unitv}[1]{\hat{\bm{#1}}}

\DeclareMathOperator{\sinc}{sinc}

\definecolor{myOrange}{RGB}{197,90,17}
\newif\ifmarked
\newif\ifmarkedd
\markedfalse

\markeddfalse

\newcommand{\annotate}[1]{%
    \ifmarked
	\textcolor{myOrange}{#1}%
    \else
	#1%
    \fi
}

\newenvironment{annotateblock}
	{\ifmarked\begingroup\color{myOrange}\fi}
	{\ifmarked\endgroup\fi}

\newcommand{\annotatebis}[1]{%
    \ifmarkedd
	\textcolor{myOrange}{#1}%
    \else
	#1%
    \fi
}

\abstract{Metasurfaces provide a compact, flexible, and reliable solution for controlling the wavefront of light. In imaging systems, micro-lens arrays are integrated with pixel matrices to reduce optical crosstalk, enhance photon collection efficiency, and improve spatial resolution. However, as the aperture size of the photonic devices decreases,  fundamental limitations associated with diffraction emerge. Here, we theoretically analyze and experimentally demonstrate that these constraints also affect the performance of small functionalized apertures, including metasurfaces and metalenses, emphasizing the increasing impact of diffraction at small pixel sizes. Despite their design versatility, our findings reveal the necessity of accounting for fundamental diffraction properties to optimize the performance of miniature optical metasurfaces.}

\begin{document}

\title[Diffraction-limited operation of micro-metalenses]{Diffraction-limited operation of micro-metalenses: fundamental bounds and designed rules for pixel integration}

\author[1]{\fnm{Nicolas}  \sur{Kossowski} \email{nicolas.kossowski@crhea.cnrs.fr}}
\author[1]{ \fnm{Christina} \sur{Kyrou}}
\author[1]{ \fnm{Rémi} \sur{Colom}}
\author[1]{\fnm{Pierre-Marie} \sur{Coulon}}
\author[1]{\fnm{Virginie} \sur{Br\"{a}ndli}}
\author[2]{\fnm{Jean-Luc} \sur{Reverchon}}
\author[1]{\fnm{Samira} \sur{Khadir}}
\author[1,3]{\fnm{Patrice} \sur{Genevet} \email{patrice.genevet@mines.edu}}

\affil[1]{\orgdiv{Université Côte d'Azur}, \orgname{CNRS, CRHEA}, \orgaddress{\street{Rue Bernard Gregory}, \city{Valbonne}, \postcode{06560}, \country{France}}}
\affil[2]{\orgdiv{III-V Lab}, \orgaddress{\street{1 Avenue Augustin Fresnel}, \city{Palaiseau}, \postcode{91767}, \country{France}}}
\affil[3]{\orgdiv{Physics department}, \orgname{Colorado School of Mines}, \orgaddress{\street{1523 Illinois Street}, \city{Golden, CO}, \postcode{80401}, \country{USA}}}

\maketitle

\section{Introduction}
Metasurfaces (MSs) consist of an arrangement of subwavelength nanostructures whose geometries and materials are carefully engineered to produce a specific response in phase and/or polarization \cite{yu_light_2011, genevet_recent_2017, yu_flat_2014}. Their operating principle relies on the collective optical response of an ensemble of nanostructures to produce user-defined optical functions. As this disruptive optical technology advances toward industrial applications, a new set of challenges associated with its integration into specific applications is emerging \cite{yoon_recent_2021}. Despite the enormous amount of recent research on MSs, studies aimed at identifying their physical limits remain elusive. Intriguingly, we observed that small MSs are not operating as expected when their diameters reach the order of a few wavelengths or less \cite{jia_focal_2018,gao_analysis_2012}. This behavior, which might at first appear as unexpected, is of particular interest in the context of vertical integration of MS in imaging systems to improve the performance of optical sensors. The vertical integration of MSs with semiconductor technologies, successfully implemented in the context of controlled laser emission \cite{xie_metasurface-integrated_2020,ni_spin-decoupling_2022, wen_vcsels_2021, wang_-chip_2021, zheng_-chip_2025}, might also bring additional industrial \annotate{prospects in light field imaging \cite{lin_achromatic_2019, fan_trilobite-inspired_2022}} and LiDAR \cite{wang_advances_2024, juliano_martins_metasurface-enhanced_2022, majorel_bio-inspired_2024}, assuming that MSs are properly designed.

Image sensor technology has benefited from the reduction in pixel size to improve both spatial resolution with higher pixel density and signal-to-noise ratio by reducing dark current \cite{rogalski_challenges_2016}. Currently, pixel sizes are reaching a critical limit of only a few wavelengths, but further reduction in pixel size introduces two major drawbacks, electrical and optical crosstalk, that hinder the modulation transfer function of the imaging system. These issues are conventionally circumvented by integrating an injection-molded array of microlenses directly onto the sensor matrix \cite{cai_microlenses_2021,beguelin_commented_2021}. However, these solutions are bulky and incur relatively expensive assembly and packaging costs to the imaging system. Additionally, molded lenses designed to be roughly the size of the pixel are characterized by an asymmetric focusing shape with a maximum intensity occurring before the designed focal length, an effect known as the \textit{focal shift} \cite{li_focal_1981}. Recent efforts to improve the optical function of conventional microlens arrays using MS are currently being investigated, as this approach could drastically simplify the fabrication process and offer new imaging functionalities, including spectral and polarization selectivity. However, the integration of these small functionalized optical apertures on top of small pixels also comes with fundamental limitations. 

Here, we propose identifying these fundamental limits, focusing particularly on the design of small metalenses, to clarify the different regimes of operation. We use a theoretical model previously proposed for small diffracting apertures, which is based on vectorial diffraction theory \cite{bouwkamp_diffraction_1954}, to account for planar apertures with spatially varying optical responses. We show that the domain of validity is bounded between (i) a hard limit defined by the MS size and (ii) a soft limit defined by the MS nanostructuration, i.e., the periodicity of the metasurface building blocks. First, we introduce the notion of \textit{focal shift} and, in the following section, we present the formalism indicating the strong expected impact of the focal shifts for sufficiently small devices. Finally, we present the experimental results obtained using \ce{GaN} metalens of various numerical apertures (NA), which validate our theoretical predictions.

\section{Results}

\begin{figure*}[!ht]
    \centering
    \includegraphics[width=\textwidth]{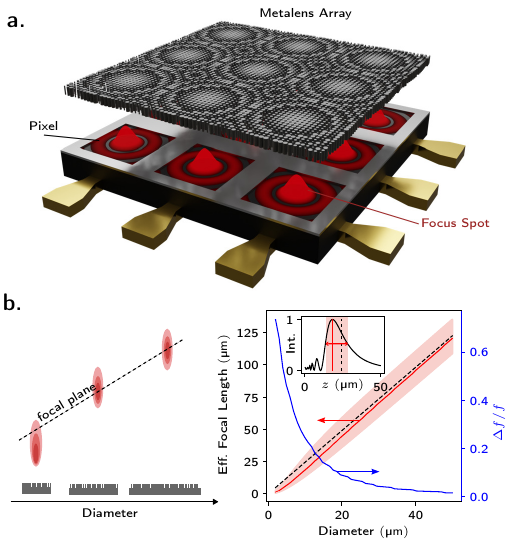}
    \caption{(a) Artistic representation of the integration of a metalens array with a matrix of pixels (b) Numerical simulation results summarizing the variation of the focal distance of metalenses designed with a fixed numerical aperture of $\text{NA}=0.2$ as a function of their diameters. The red curve represents the effective focal distance, and the black dotted line represents the designed focal length. The red area represents the effective full width at half maximum (FWHM) of the spot in the $z$-direction, as indicated in the insert. Finally, the blue curve represents the focal shift error $\Delta f/f = \abs{f_{eff}-f}/f$ as a function of the metalens diameter. The simulations were carried for a wavelength of $\SI{617}{\nano\metre}$.}
    \label{fig:fig1}
\end{figure*}

Fig.~\ref{fig:fig1}(a) illustrates the concept of vertical integration of a metalens array onto a sensor. All metalenses depicted in the schematic are identical and designed to focus light along their optical axis, i.e. at the center of their respective pixels. We begin our investigation by numerically calculating, using diffraction theory, the focal distance of a metalens as a function of its diameter with a fixed numerical aperture (NA) given by,
 \begin{equation}
    \text{NA} = n\sin\left[\arctan\left(\frac{a}{f}\right)\right],
    \label{eq:na}
 \end{equation}
where $a$ is the radius of the lens, $f$ its focal distance, and $n$ the refractive index of the medium. Here, we consider that only a layer of air ($n=1$) separates the metalens array from the absorbing pixels. From this expression, we see that decreasing the radius of a metalens with fixed NA should result in a proportional decrease in its focal length. However, in practice, when the lens reaches a critical aperture size, this simple law breaks down, as illustrated in the numerical calculations presented in Fig.~\ref{fig:fig1}(b). 
Instead, we observe that for a diameter smaller than $\SI{20}{\micro\metre}$ and a wavelength of \SI{617}{\nano\metre}, the maximum intensity, or effective focal length, occurs at a shorter distance than the designed focal length. At the same time, the spot becomes highly asymmetric along the propagation direction, with the maximum intensity shifted closer toward the metalens. This phenomenon has been previously reported as the so-called \textit{focal shift} \cite{li_focal_1981}. The origin of the focal shift is related to the diffraction that occurs when the size of the metalens aperture $a$ approaches the operation wavelength $\lambda$.

\subsection{General Theory of Diffraction By Wavelength-scale Apertures}

We consider the transmission of a plane wave from the $z$-negative region of space through a metasurface of aperture $A$, with radius $a>\lambda$, and located in the plane $z=0$ (see Fig.~\ref{fig:fig2}(a)). We assume that the electromagnetic fields at $z=0$ are zero everywhere outside of the aperture, and that inside the aperture, \annotate{it is noted $(\bm{E}_a, \bm{H}_a)$}. \annotate{The presence of field discontinuities in the aperture plane results in wide-angle beam deflection, which can be rigorously treated mathematically using a vectorial formulation of the diffraction problem. Its solution is given by the Stratton-Chu integral} \cite{bouwkamp_diffraction_1954, stratton_diffraction_1939}. Adopting the Gaussian unit system, the diffracted electric field distribution after the aperture $\bm{E}_d$ reads as
\begin{equation}
    \begin{aligned}
        \bm{E}_d(\bm{r}) = -\frac{1}{4\pi} \iint_A \bigg\{ & ik\left[\unitv{n}\times\bm{H}_a(\bm{r}^\prime)\right]G(\bm{r},\bm{r}^\prime) \\
        & + \left[\unitv{n}\times\bm{E}_a(\bm{r}^\prime)\right]\times\nabla G(\bm{r},\bm{r}^\prime) \\
        & + \left[\unitv{n}\cdot\bm{E}_a(\bm{r}^\prime)\right]\nabla G(\bm{r},\bm{r}^\prime)\bigg\}d \bm{r}^\prime\\
        & - \frac{1}{4\pi k} \oint_C \nabla G(\bm{r},\bm{r}^\prime) [\bm{H}_a(\bm{r}^\prime)\cdot \unitv{l}] d \bm{r}^\prime,
    \end{aligned}
    \label{eq:equivTh}
\end{equation}
with $G(\bm{r}, \bm{r}^\prime) = \exp(ik\abs{\bm{r} - \bm{r}^\prime})/\abs{\bm{r} - \bm{r}^\prime}$ the Green's function. The unit vector $\unitv{n}$ points towards the negative space, i.e. $\unitv{n}=-\unitv{z}$, and the unit vector $\unitv{l}$, tangent to the aperture, is defined in the positive direction with regards to the normal $\unitv{n}$. The surface integral on the right-hand side of \annotate{Eq.~}\eqref{eq:equivTh} describes the field radiated by sources of electric and magnetic currents, as well as charge distribution, inside the aperture. Under certain assumptions, this term alone can be reduced to the well-known expressions of Fresnel and Fraunh\"{o}fer diffraction theory \cite{born_principle_1999}. However, the field described only by the surface integral does not properly satisfy Maxwell's equations at the location of the aperture. To derive \annotate{Eq.~}\eqref{eq:equivTh}, we assume that the fields and their derivatives should be continuous on the screen and aperture. This is not the case as $\bm{E}_a(a^+)=0$ and $\bm{E}_a(a^-)\neq 0$ while increasing the radius $r$ to pass the contour $C$ delimiting the aperture. The contour integral in \annotate{Eq.~}\eqref{eq:equivTh} resolves this inconsistency by considering Maxwell's continuity equations of the fields on the screen through the contour $C$ of the aperture \cite{kottler_elektromagnetische_1923}.

\begin{annotateblock}
    In order to compute the diffracted field $\bm{E}_d$ by the metasurface, we need to determine the electromagnetic field $(\bm{E}_a, \bm{H}_a)$, inside the aperture. For a metasurface, with a spatially varying phase profile $\varphi (\bm{r})$ and transmission coefficient $t(\bm{r})$, the wavefront of the transmitted light thus follows the generalized law of refraction~\cite{yu_light_2011, aieta_out--plane_2012},
    \begin{equation}
        \bm{k}_{a,\perp} = \bm{k}_{i,\perp} + \bm{D}\varphi,
        \label{eq:generalized_law_refraction}
    \end{equation}
    with $\bm{k}_{i,\perp}$ and $\bm{k}_{a,\perp}$ the transverse wavevector of the incident wave and of the transmitted wave, respectively and $\bm{D}\varphi$ a vector composed of the derivatives of the phase profile. Consequently, the transmitted wave is not necessarily a plane wave; in the case of metalenses the wavefront is spherical. Moreover, according to Maxwell's equations, the electric field is locally transverse to the direction of propagation, therefore the curvature of the wavefront leads to a complex distribution of polarization of the transmitted wave. In the particular case of spherical metalenses, we can determine this polarization pattern and thus express the electromagnetic field inside the aperture, cf. supplementary note~\ref{sm-sec:vectorial_diff}.
\end{annotateblock}

\subsection{Origin of the Focal Shift} \label{sec:circAperture}

\begin{figure*}[!ht]
    \centering
    \includegraphics[width=\textwidth]{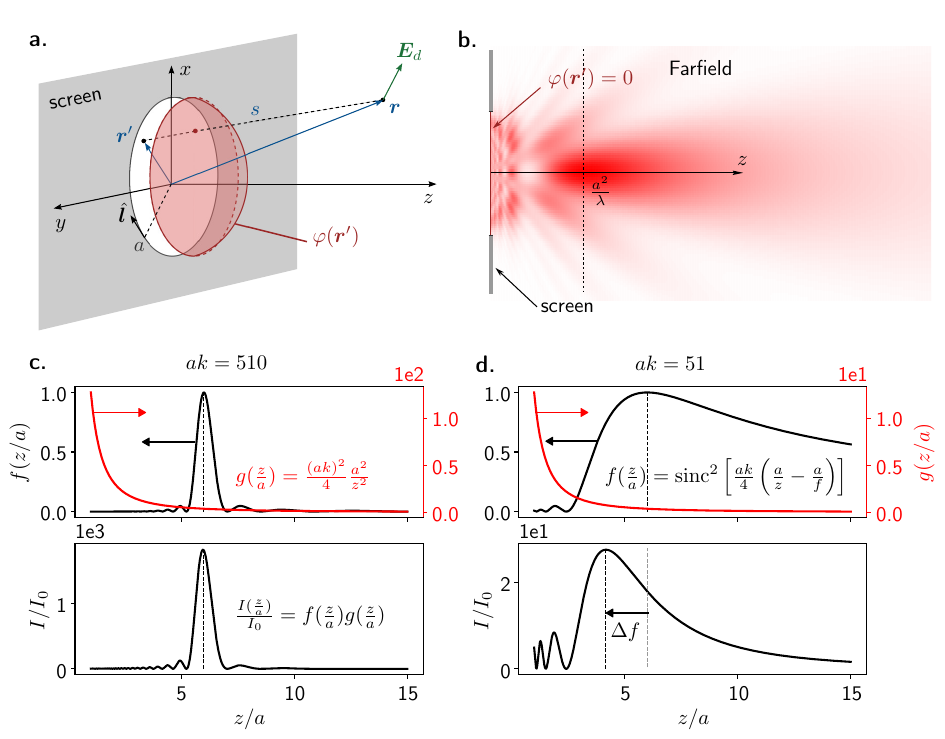}
    \caption{(a) Schematic of a circular transmitting aperture in an opaque screen capable of imparting an additional transmission phase delay function $\phi(\bm{r}^\prime)$ on the incoming light, as illustrated with a red overlaying profile. (b) Schematic of the diffraction pattern for a plain aperture, $\varphi(\bm{r})=0$. The vertical black dashed line represents the focal plane of the aperture, approximately located at $a^2/\lambda$. (c) and (d) show the normalized intensity (bottom panels) along the $z$-axis for two metalenses with fixed focal distance of $f=6a$ and with $ak =510$ (left panel) and $ak=51$ (right panel), respectively. Top panels show the focusing function $f(x) = \sinc^2\left[ \frac{ak}{4}\left( \frac{1}{x}-\frac{a}{f} \right) \right]$ in black with the left ordinate and intensity decay $g(x) = \frac{(ak)^2}{4} \frac{1}{x}$ in red with the right red ordinate. For $ak = 510$, there is no focal shift, while for $ak = 51$, a focal shift occurs.}
    \label{fig:fig2}
\end{figure*}

\begin{annotateblock}
    To describe the focal shift, it is sufficient to look at the diffracted electric field $\bm{E}_d$ on the optical axis $z$. By symmetry, for a metalens, only the component of the electric field parallel to the incident polarization remains, since the orthogonal components cancel on the optical axis, cf. supplementary note~\ref{sm-sec:vectorial_diff}. Therefore, Eq.~\eqref{eq:equivTh} can be treated analytically. In the following, we assume that the metasurface has unitary and uniform transmission, $t(\bm{r}) = 1$.
\end{annotateblock}

We first consider the case of a plain aperture, represented in Fig.~\ref{fig:fig2}(b) with homogeneous response. We are interested in the field at a large distance from the aperture. In this case, the intensity along the $z$-axis at $z > a$, derived in supplementary note~\ref{sm-sec:ap_circAperture}, is given by
\begin{equation}
    I(z) = I_0 \frac{a^4k^2}{4z^2} \sinc^2\left[ \frac{a^2k}{4z} \right].
    \label{eq:int_circAperture}
\end{equation}
A map of the intensity distribution in the $xz$-plane, obtained after solving \annotate{Eq.~}\eqref{eq:equivTh}, is plotted in Fig.~\ref{fig:fig2}(b). \annotate{Equation~\eqref{eq:int_circAperture} illustrates a known result: even in the absence of a phase profile, a circular aperture, i.e. a pinhole,} focuses light at a certain distance $f_{\text{ap}}$. A detailed study of the intensity function $I(z)$, in supplementary note~\ref{sm-sec:ap_circAperture}, shows that the maximum of intensity is approximately obtained for
\begin{equation}
    f_{\text{ap}} \approx \frac{a^2}{\lambda}.
    \label{eq:z1}
\end{equation}
The asymmetry in the shape of the focal spot arises from the dependence of the argument of $\sinc$ on the inverse distance $1/z$.

\annotate{Now, we consider the case of a metasurface placed in the aperture and with an hyperbolic phase profile $\varphi(\bm{r}^\prime)$ given by the equation,}
\begin{equation}
    \varphi(\bm{r}^\prime) = k\left(f - \abs{\bm{r}^\prime-f\unitv{z}}\right).
    \label{eq:ph_lens}
\end{equation}
\annotatebis{Since, we are interested in the region of focal shift, and therefore, we can simplify Eq.~\eqref{eq:equivTh}.} \annotate{Similarly to the case of the plain aperture} in Eq.~\eqref{eq:int_circAperture}, we derive, \annotate{in the supplementary note~\ref{sm-sec:circLens}}, the intensity distribution along the optical axis for $z > a$, \annotatebis{for a metalens with numerical aperture lower than $0.7$,}
\begin{equation}
    I(z) = I_0 \frac{a^4k^2}{4z^2} \sinc^2\left[ \frac{ak}{4}\left( \frac{a}{z}-\frac{a}{f} \right) \right].
    \label{eq:int_circLens}
\end{equation}
Equation~\eqref{eq:int_circLens} is analogous to \annotate{Eq.~}\eqref{eq:int_circAperture} except for an additional term in the $\sinc$ related to the focal distance of the metalens. The effective focal length is given by the location of the maximum intensity, which results from the product of the two terms: (i) the $\sinc^2$ function representing the z-focusing profile and (ii) the intensity decay, which is proportional to $1/z^2$. When considering metalenses with a large focal length compared to their aperture size $a$, $f \gg a$, the intensity distribution converges to that of the homogeneous aperture. This is not entirely surprising as $f \gg a$ implies that the profile at the aperture plane is converging to a constant spatial phase value. Conversely, if the focal distance is smaller than the aperture of the metalens, $f \ll a$, the metalens will focus when the $\sinc$ is maximal, at $z = f$. In the intermediate regime, however, the maximum of intensity \annotate{is not obtained} at $z = f$ even though the $\sinc$ \annotate{is maximal}. Indeed, the intensity is quenched by the term $1/z^2$. The effective focal distance will depend on the relative value between the aperture size $a$, the focal distance $f$, and the operating wavelength $\lambda$. For example, fixing the focal distance $f = 6a$, on one hand, for a large quantity $ak = 510$, i.e. for a large aperture compared to the wavelength, the width of the $\sinc$ is thin and thus the effect of intensity decrease will be negligible, cf. Fig.~\ref{fig:fig2}(c). On the other hand, for smaller $ak = 51$, the $\sinc$ term is large and asymmetric, leading to a strong focal shift at a shorter distance than the expected focal length, cf. Fig.~\ref{fig:fig2}(d). 

The interplay between the three fundamental properties of a lens, i.e. $a$ the characteristic size of the lens, $f$ the focal length, and  $\lambda$ the operating wavelength, is summarized in the Fresnel number \cite{li_dependence_1982},
\begin{equation}
    N = \frac{a^2}{\lambda f}.
    \label{eq:fresnelNb}
\end{equation}
An expression of \annotate{Eq.~}\eqref{eq:int_circLens} in terms of the Fresnel number is given in the \annotate{supplementary note~\ref{sm-sec:circLens}}. In the literature, the Fresnel number is often referred to as the number of half-wavelengths in the wavefront at the edge of the aperture \cite{wang_far-zone_1995}. This derivation can also be obtained for a small numerical aperture by performing a Taylor expansion of expression \annotate{Eq.~}\eqref{eq:ph_lens}.
Our analysis in \annotate{supplementary note} \ref{sm-sec:ap_circAperture} also suggests that the Fresnel number can be interpreted as the ratio of the focal distance of the plain aperture $f_{\text{ap}}$ to the characteristic length of the optical function achieved inside the aperture $f_{\text{func}}$,
\begin{equation}
    N = \frac{f_{\text{ap}}}{f_{\text{func}}}.
    \label{eq:fresnelRatio}
\end{equation}
For $N>1$, the optical function is achieved before the focal spot of the aperture, meaning that the metalens primarily acts as a lens. Conversely, for $N<1$ the optical function is achieved beyond the focal spot of the aperture, so the metalens primarily acts as a simple aperture. In other words, \annotate {the aperture selects only a small part of the overall focused phase information, i.e., reducing the overall phase curvature to what is in the aperture}. We define the \textit{diffraction-limited operation regime of micro-metalenses}, the regime where the diffraction of the aperture itself compromises the metasurface's optical function.

To visually illustrate the evolution of the Fresnel number as a function of the three fundamental parameters, we plotted in Fig.~\ref{fig:fig4}(a) a mapping of the Fresnel number as a function of the number of wavelengths inside the aperture $a/\lambda$ and also indicating the numerical aperture. In the case of microlenses, for example $a = 10\lambda$, the Fresnel number is less than $10$ for numerical apertures smaller than approximately $0.75$, suggesting a strong focal shift. To avoid focal shift, then, metalenses would require a larger NA. Additionally, we can also identify a minimal numerical aperture, $\text{NA} < 0.1$, for which we would have $N > 1$ and thus, the micro-metalenses would not behave significantly differently from an aperture. From Fig.~\ref{fig:fig4}(a), we can also see that the focal shift occurs mainly with small metalenses. For $a = 100\lambda$, the Fresnel number exceeds $10$ for nearly any numerical aperture above $0.1$.

\subsection{Limitation of the phase sampling}

\begin{figure*}[ht!]
    \centering
    \includegraphics[width=\textwidth]{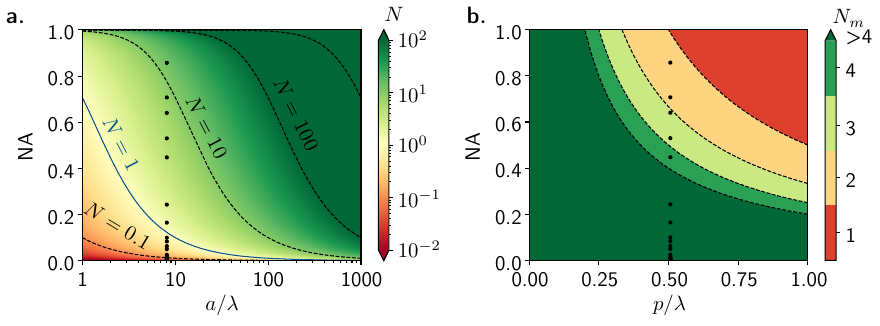}
    \caption{(a) Fresnel number $N$ as a function of the normalized aperture size $a/\lambda$ and the numerical aperture. The blue line represents $N=1$, the green colors $N>1$, and the orange colors $N<1$. The dashed black lines represent power of $10$. (b) Number of elements per Fresnel zone $N_m$ as a function of the period of the metasurface and the numerical aperture. The black dot represents the designed metalenses fabricated and characterized with $p = \SI{312.5}{\nano\metre}$ and $a = \SI{10}{\micro\metre}$. In both figures the wavelength is taken equal to \SI{617}{\nano\metre}.}
    \label{fig:fig4}
\end{figure*}

So far, the phase profiles imparted by the metasurface have been considered continuous. In practice, phase shifts are induced by discrete meta-atoms of finite size. For adequate sampling of the phase profile, we require at least $N_m \geq 2$ meta-atoms per Fresnel zone. Following this rule, we obtain, in supplementary note~\ref{sm-sec:sampling}, a second limit for metalenses, 
\begin{equation}
    \text{NA} \frac{p}{\lambda} < \frac{1}{N_m},
    \label{eq:na_p}
\end{equation}
with $p$ the period separating two adjacent metasurface building blocks. This is a soft limit, and achieving higher discretizations depends on the capacity of our technological platform to fabricate sufficiently small meta-atoms to meet the condition~\eqref{eq:na_p}. According to the Shannon-Nyquist sampling theorem, to properly sample the phase profile, only two meta-atoms per Fresnel zone are sufficient. However, for such sampling, the Strehl ratio of a metalens was estimated to reach only a value of $0.4$ \cite{aieta_aberrations_2013}. Achieving Strehl ratios associated with diffraction-limited performance, beyond $0.8$, requires at least four elements per Fresnel zone, $p/\lambda < 0.25$. For phase profile~\eqref{eq:ph_lens} with fixed focal length, the size of a Fresnel zone decreases and converges toward $\lambda$ as we move further away from the device center. This results from the linear asymptotic behavior of the phase profile at large distance $\bm{r}^\prime$, bending the light at \SI{90}{\degree}. This limit is obtained only for large distances from the metalens center, that is, for diameters that easily exceed the metalens aperture size. For micro-metalenses, which are designed with extremely small diameters, barely equal to a few wavelengths, this distance is quickly reached, resulting in poor sampling of the phase profile.

Similarly to the previous section, we can visually represent \annotate{Eq.~}\eqref{eq:na_p} by plotting the number of elements per Fresnel zone $N_m$ in Fig.~\ref{fig:fig4}(b). The plot shows that, to achieve optimal performance for optical micro-metalenses, one should avoid the diffractive regime by increasing the numerical aperture, but it also requires designing structures with smaller critical dimensions.

\subsection{Fabrication and Experimental Characterization}

\begin{figure*}[ht!]
    \centering
    \includegraphics[width=\textwidth]{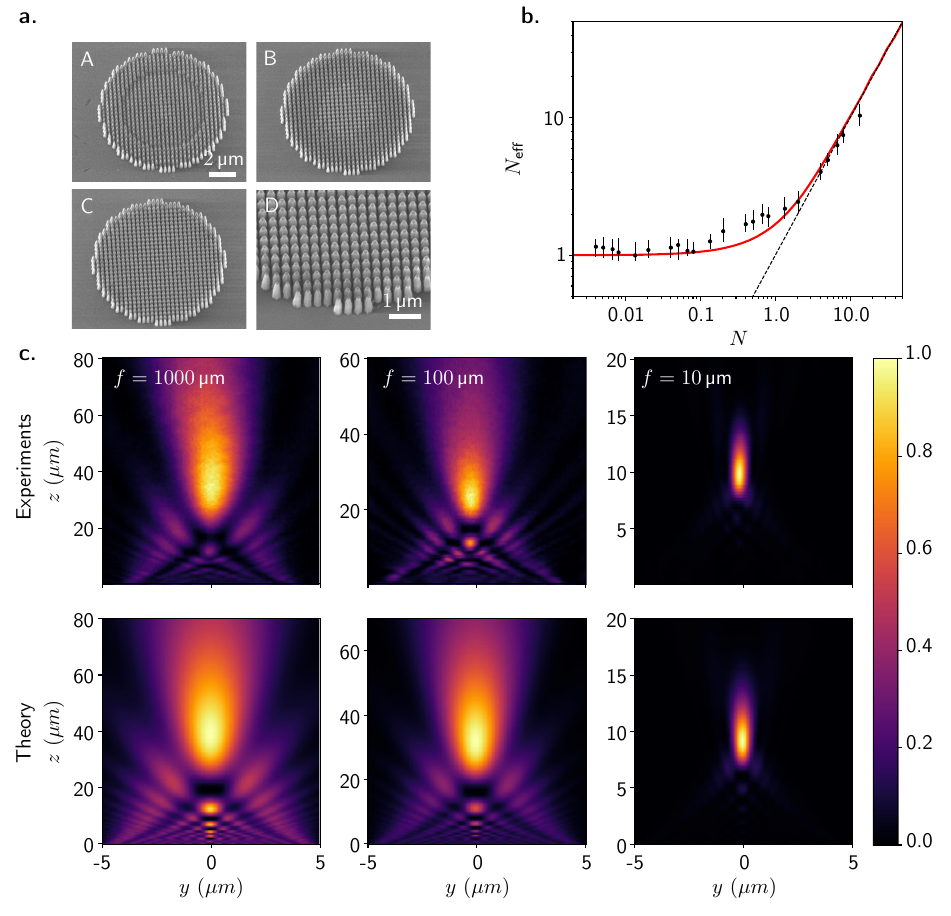}
    \caption{(a) Scanning electron micrographs of the \ce{GaN} metalenses with focal length  (A) $f=\SI{8}{\micro\metre}$, (B) $f=\SI{500}{\micro\metre}$ and (C-D) $f=\SI{3000}{\micro\metre}$. (D) The scale bar represents $\SI{2}{\micro\metre}$ for (A-C) and $\SI{1}{\micro\metre}$ for (D). (b) Evolution of the effective Fresnel number, $N_{\text{eff}} = a^2/\lambda f_{\text{eff}}$ and expected Fresnel number $N = a^2/\lambda f$. The red curve corresponds to the simulation results. (c) Comparison between numerical (bot panels) and experimental (top panels) intensities for metalenses of diameter $\SI{10}{\micro\metre}$ and focal length $\SI{1000}{\micro\metre}$ (left panels), $\SI{100}{\micro\metre}$ (middle panels) and $\SI{10}{\micro\metre}$ (right panels). \annotate{The numerical simulations have been done using numerical integration in Eq.~\eqref{eq:equivTh}.}}
    \label{fig:fig3}
\end{figure*}

To validate our approach, we experimentally characterized an array of \SI{10}{\micro\metre}-wide metalenses, see Fig.~\ref{fig:fig3}(a) \annotate{and method~\ref{sec:method_fab}}. The fabricated meta-atoms are arranged on a square grid with period $\SI{312.5}{\nano\metre}$ designed to operate at wavelength $\SI{617}{\nano\metre}$. Therefore, in Fig.~\ref{fig:fig3}(a) we are located at $a/\lambda \approx 8$ and in Fig.~\ref{fig:fig4}(b) at $p/\lambda \approx 0.5$. We realized an array of $23$ metalenses \annotate{with fix diameters of \SI{10}{\micro\metre} and} with varying focal length ranging from \SI{10}{\milli\metre} to \SI{3}{\micro\metre}. Metalenses with focal length smaller than $\SI{2.4}{\micro\metre}$ would have a numerical aperture higher than the numerical aperture of our objective ($\annotate{\text{NA} = } 0.9$). Each metalens was characterized by a $z$-scan setup to determine its \textit{effective} focal length.

The experimental z-scan intensity mappings are presented in Fig.~\ref{fig:fig3}(c) for designed focal length of \SI{1}{\milli\metre}, \SI{100}{\micro\metre} and \SI{10}{\micro\metre} and \annotate{compared with numerical simulation where we solved Eq.~\eqref{eq:equivTh}. The z-scan characterization setup is described in method~\ref{sec:method_charac}}. Our results show acceptable agreement. The effective Fresnel number of the fabricated metalenses is plotted in Fig.~\ref{fig:fig3}(b). The error bars represent the accumulated uncertainty in determining the metasurface plane for each metalens, as well as the maximum intensity, which is evaluated around 80\% of the maximum intensity. For low Fresnel numbers, $N < 0.1$, the metalenses behave as simple apertures with effective Fresnel numbers approaching $1$. For Fresnel numbers comprised between $0.1$ and $1$, slight deviations of experimental measurements from theoretical results are attributed to our fabrication processes, which lead to imperfect optical performance. For Fresnel numbers comprised between $1$ and $10$, experimental results agree with the theory, following the expected linear dependency \annotate{Eq.}~\eqref{eq:na}.

\section{Discussion}
Replacing refractive microlenses with vertically integrated metalenses arrays offers an innovative technological alternative to improve spatial resolution and reduce the integration issues. While there are fundamental optical and fabrication consideration, these challenges present opportunities for creative integration and design solutions. Designing such a compact metalens array involves finding an optimal compromise between the metalens numerical aperture and its structural parameters. 

For a fixed Fresnel number, \annotate{Eq.~}\eqref{eq:fresnelNb} predicts that decreasing the metalens diameter by a factor $\alpha$ results in a focal length reduction by $\alpha^2$. In most large metalenses, meaning their diameter greatly exceeds the operating wavelength, accuracy in the focal point is not an issue. However, designing small microlenses, such as those with dimensions below \SI{10}{\micro \metre} used in camera pixels to enhance detectivity, requires carefully selecting the right NA and design assembly. We demonstrated that in low-NA microlenses associated with low Fresnel numbers, diffraction effects dominate, limiting the lens performance. With high-NA microlenses, the period of the \annotate{meta-atoms} forming the metasurface should satisfy the condition in Eq.~\eqref{eq:na_p}, otherwise, the sampling period will be insufficient to ensure optimal diffraction-limited performances. Additionally, metalens array integration generally requires dealing with the chip substrate, which incidentally could modify the effective wavelength and the Fresnel number, requiring further adjustments in design, including sampling period and focal distance.  

We identified four distinct operational regimes: (i) the diffraction regime, where diffraction dominates and the metalens function as a simple aperture; (ii) the hybrid-diffractive regime, where the lensing effect is present but strongly affected by the diffraction; (iii) the metasurface regime, where the metalens operates effectively with proper wavefront control; and (iv) the metasurface sampling-limited regime, where high NA wavefront engineering requires improving the technical processes to address higher aspect ratio fabrication.

Maximizing light collection on a conventional 2D square pixel array is more effective with square lenses rather than circular lenses discussed here. Since the NA is defined for optical lenses with revolution symmetry, the Fresnel number of a square metalens, $N_{\text{sq}}$, falls within the range $N<N_{\text{sq}}<2N$, which provides a useful information for design consideration, according to the operation regimes outlined above. 

\begin{annotateblock}
    Following this discussion, a natural question emerges: when is it better to use a metasurface than a conventional aperture? To answer this question, we can look at the focusing efficiency, defined by the amount of power going through one pixel, divided by the amount of power transmitted through the metasurface, see supplementary note~\ref{sm-sec:focal_efficiency}. For Fresnel number larger than $1$, metasurfaces act better than an actual aperture. However, with a non-optimized approach, such as a using a fixed library of elements on fix grid, gain in terms of focusing efficiency could be only few percent. Since, for micro-metalenses, the number of different meta-atoms is only about few dozens, we believe optimization methods \cite{elsawy_multiobjective_2021} could be particularly interesting for such applications. 
\end{annotateblock}

In conclusion, we explore the fundamental limitations of miniaturized metalenses, with sizes only a few times the operation wavelength. Using diffraction theory, we modeled a finite-sized ideal metasurface and analyzed how the metalens' dimensions impact its optical functionality. Our findings reveal two key operational regimes based on the device's Fresnel number. Operating metalenses in the diffractive regime (low Fresnel number) presents challenges due to the alteration of the wavefront by the diffraction resulting from the metasurface aperture, making the aperture optical functionalization ineffective. Instead, for arrays made of very small metalenses, $a\lesssim 10\lambda$ when working at visible/IR wavelengths, achieving a high numerical aperture and thus a short focal length is essential for optimal focusing properties. This work provides new insights into metalens integrations and discusses the design opportunities available for compact imaging systems, particularly when the photonic devices are directly fabricated onto the pixel matrices. Our results indicate that, depending on parameters, metasurfaces could be a versatile platform for enhancing imaging capabilities, paving the way for next-generation sensor technology.

\section{Methods}
\subsection{Metalenses Fabrication}\label{sec:method_fab}
The metalenses have been fabricated by employing \SI{1}{\micro\metre}-thick \ce{GaN} epitaxially grown by metal organic vapor phase epitaxy on a double side polish sapphire substrate. The metalenses were patterned by exposing a double layer of PMMA resist using a focused electron beam, followed by development, and deposition of a \ce{Ni} mask of $\SI{70}{\nano\metre}$. The cylindrical $\SI{1}{\micro\metre}$-thick \ce{GaN} nanopillars have finally been etched on the substrate by employing an inductively coupled plasma reactive ion etching process based on \ce{Cl_2} and \ce{BCl_3} followed by the chemical removal of the metallic mask in the Piranha solution.

\subsection{Z-scan Characterization}\label{sec:method_charac}
A collimated LED light source was used to illuminate the metalenses. The sample holder was mounted on a motorized translation stage controlled by a closed-loop picomotor controller. The size of the translation stage step is \SI{50}{\nano\metre}. \annotate{An objective, with $\times 100$ magnification and $0.9$ numerical aperture,} is used to image each metalens separately onto a camera. The plane of the metalens was visually determined by the camera. Due to the uncertainty in determining this plane, we identified the translation range for which we could consider the image on camera to be in the plane of the metasurface and estimated this uncertainty to be within the range $\pm\SI{400}{\nano\metre}$.

\backmatter
\bmhead{Data Availability}
Data that support the plots in this paper and other findings of this study are available from the corresponding authors upon reasonable request.

\bmhead{Code Availability}
The codes for numerical simulations of this study are available from the corresponding authors upon reasonable request.

\bmhead{Acknowledgements} 
PG, NK, JLR, SK acknowledge financial support from the French National Research Agency ANR Project Millesime 2 (ANR-21-ASTR-0014). SK acknowledges financial support from the European Defense fund EDF project Mini-Bot (EDF-2021-OPEN-R-SME-2). RC was supported by the French government through the France 2030 investment plan managed by the National Research Agency (ANR), as part of the Initiative of Excellence Université Côte d’Azur
under reference number ANR-15-IDEX-01. PG and CK acknowledge financial support from the French National Research Agency ANR Project SWEET (ANR-22-CE24-00005).

\bmhead{Author Contributions} PG and JLR conceived this study. NK worked on the theoretical part, the numerical simulations and optical characterization. PMC grew the \ce{GaN} template used for the fabrication of the metalenses. CK did the fabrication processes. VB helped with the electron-beam lithography as well as for SEM characterization. RC, JLR, SK and PG reviewed and edited the manuscript. All authors edited the manuscript at several stages and contributed to the references.

\bmhead{Competing Interests}
The authors declare no competing interests.

\bibliography{ref}

\begin{thebibliography}{10}
\expandafter\ifx\csname url\endcsname\relax
  \def\url#1{\burl{#1}}\fi
\expandafter\ifx\csname urlprefix\endcsname\relax\def\urlprefix{URL }\fi
\providecommand{\bibinfo}[2]{#2}
\providecommand{\eprint}[2][]{\url{#2}}
\providecommand{\doi}[1]{\url{https://doi.org/#1}}
\bibcommenthead

\bibitem{yu_light_2011}
\bibinfo{author}{Yu, N.} \emph{et~al.}
\newblock \bibinfo{title}{Light {{Propagation}} with {{Phase Discontinuities}}: {{Generalized Laws}} of {{Reflection}} and {{Refraction}}}.
\newblock \emph{\bibinfo{journal}{Science}} \textbf{\bibinfo{volume}{334}}, \bibinfo{pages}{333--337} (\bibinfo{year}{2011}).

\bibitem{genevet_recent_2017}
\bibinfo{author}{Genevet, P.}, \bibinfo{author}{Capasso, F.}, \bibinfo{author}{Aieta, F.}, \bibinfo{author}{Khorasaninejad, M.} \& \bibinfo{author}{Devlin, R.}
\newblock \bibinfo{title}{Recent advances in planar optics: From plasmonic to dielectric metasurfaces}.
\newblock \emph{\bibinfo{journal}{Optica}} \textbf{\bibinfo{volume}{4}}, \bibinfo{pages}{139--152} (\bibinfo{year}{2017}).

\bibitem{yu_flat_2014}
\bibinfo{author}{Yu, N.} \& \bibinfo{author}{Capasso, F.}
\newblock \bibinfo{title}{Flat optics with designer metasurfaces}.
\newblock \emph{\bibinfo{journal}{Nature Materials}} \textbf{\bibinfo{volume}{13}}, \bibinfo{pages}{139--150} (\bibinfo{year}{2014}).

\bibitem{yoon_recent_2021}
\bibinfo{author}{Yoon, G.}, \bibinfo{author}{Tanaka, T.}, \bibinfo{author}{Zentgraf, T.} \& \bibinfo{author}{Rho, J.}
\newblock \bibinfo{title}{Recent progress on metasurfaces: Applications and fabrication}.
\newblock \emph{\bibinfo{journal}{Journal of Physics D: Applied Physics}} \textbf{\bibinfo{volume}{54}}, \bibinfo{pages}{383002} (\bibinfo{year}{2021}).

\bibitem{jia_focal_2018}
\bibinfo{author}{Jia, Y.}
\newblock \bibinfo{title}{Focal shift in metasurface based lenses}.
\newblock \emph{\bibinfo{journal}{Optics Express}} \textbf{\bibinfo{volume}{26}}, \bibinfo{pages}{8001--8015} (\bibinfo{year}{2018}).

\bibitem{gao_analysis_2012}
\bibinfo{author}{Gao, Y.} \emph{et~al.}
\newblock \bibinfo{title}{Analysis of focal-shift effect in planar metallic nanoslit lenses}.
\newblock \emph{\bibinfo{journal}{Optics Express}} \textbf{\bibinfo{volume}{20}}, \bibinfo{pages}{1320--1329} (\bibinfo{year}{2012}).

\bibitem{xie_metasurface-integrated_2020}
\bibinfo{author}{Xie, Y.-Y.} \emph{et~al.}
\newblock \bibinfo{title}{Metasurface-integrated vertical cavity surface-emitting lasers for programmable directional lasing emissions}.
\newblock \emph{\bibinfo{journal}{Nature Nanotechnology}} \textbf{\bibinfo{volume}{15}}, \bibinfo{pages}{125--130} (\bibinfo{year}{2020}).

\bibitem{ni_spin-decoupling_2022}
\bibinfo{author}{Ni, P.-N.} \emph{et~al.}
\newblock \bibinfo{title}{Spin-decoupling of vertical cavity surface-emitting lasers with complete phase modulation using on-chip integrated {{Jones}} matrix metasurfaces}.
\newblock \emph{\bibinfo{journal}{Nature Communications}} \textbf{\bibinfo{volume}{13}}, \bibinfo{pages}{7795} (\bibinfo{year}{2022}).

\bibitem{wen_vcsels_2021}
\bibinfo{author}{Wen, D.}, \bibinfo{author}{Meng, J.}, \bibinfo{author}{Cadusch, J.~J.} \& \bibinfo{author}{Crozier, K.~B.}
\newblock \bibinfo{title}{{{VCSELs}} with {{On-Facet Metasurfaces}} for {{Polarization State Generation}} and {{Detection}}}.
\newblock \emph{\bibinfo{journal}{Advanced Optical Materials}} \textbf{\bibinfo{volume}{9}}, \bibinfo{pages}{2001780} (\bibinfo{year}{2021}).

\bibitem{wang_-chip_2021}
\bibinfo{author}{Wang, Q.-H.} \emph{et~al.}
\newblock \bibinfo{title}{On-{{Chip Generation}} of {{Structured Light Based}} on {{Metasurface Optoelectronic Integration}}}.
\newblock \emph{\bibinfo{journal}{Laser \& Photonics Reviews}} \textbf{\bibinfo{volume}{15}}, \bibinfo{pages}{2000385} (\bibinfo{year}{2021}).

\bibitem{zheng_-chip_2025}
\bibinfo{author}{Zheng, C.-L.}, \bibinfo{author}{Ni, P.-N.}, \bibinfo{author}{Xie, Y.-Y.} \& \bibinfo{author}{Genevet, P.}
\newblock \bibinfo{title}{On-chip light control of semiconductor optoelectronic devices using integrated metasurfaces}.
\newblock \emph{\bibinfo{journal}{Opto-Electronic Advances}} \textbf{\bibinfo{volume}{8}}, \bibinfo{pages}{240159--26} (\bibinfo{year}{2025}).

\bibitem{lin_achromatic_2019}
\bibinfo{author}{Lin, R.~J.} \emph{et~al.}
\newblock \bibinfo{title}{Achromatic metalens array for full-colour light-field imaging}.
\newblock \emph{\bibinfo{journal}{Nature Nanotechnology}} \textbf{\bibinfo{volume}{14}}, \bibinfo{pages}{227--231} (\bibinfo{year}{2019}).

\bibitem{fan_trilobite-inspired_2022}
\bibinfo{author}{Fan, Q.} \emph{et~al.}
\newblock \bibinfo{title}{Trilobite-inspired neural nanophotonic light-field camera with extreme depth-of-field}.
\newblock \emph{\bibinfo{journal}{Nature Communications}} \textbf{\bibinfo{volume}{13}}, \bibinfo{pages}{2130} (\bibinfo{year}{2022}).

\bibitem{wang_advances_2024}
\bibinfo{author}{Wang, X.} \emph{et~al.}
\newblock \bibinfo{title}{Advances in information processing and biological imaging using flat optics}.
\newblock \emph{\bibinfo{journal}{Nature Reviews Electrical Engineering}} \textbf{\bibinfo{volume}{1}}, \bibinfo{pages}{391--411} (\bibinfo{year}{2024}).

\bibitem{juliano_martins_metasurface-enhanced_2022}
\bibinfo{author}{Juliano~Martins, R.} \emph{et~al.}
\newblock \bibinfo{title}{Metasurface-enhanced light detection and ranging technology}.
\newblock \emph{\bibinfo{journal}{Nature Communications}} \textbf{\bibinfo{volume}{13}}, \bibinfo{pages}{5724} (\bibinfo{year}{2022}).

\bibitem{majorel_bio-inspired_2024}
\bibinfo{author}{Majorel, C.} \emph{et~al.}
\newblock \bibinfo{title}{Bio-inspired flat optics for directional {{3D}} light detection and ranging}.
\newblock \emph{\bibinfo{journal}{npj Nanophotonics}} \textbf{\bibinfo{volume}{1}}, \bibinfo{pages}{1--10} (\bibinfo{year}{2024}).

\bibitem{rogalski_challenges_2016}
\bibinfo{author}{Rogalski, A.}, \bibinfo{author}{Martyniuk, P.} \& \bibinfo{author}{Kopytko, M.}
\newblock \bibinfo{title}{Challenges of small-pixel infrared detectors: A review}.
\newblock \emph{\bibinfo{journal}{Reports on Progress in Physics}} \textbf{\bibinfo{volume}{79}}, \bibinfo{pages}{046501} (\bibinfo{year}{2016}).

\bibitem{cai_microlenses_2021}
\bibinfo{author}{Cai, S.} \emph{et~al.}
\newblock \bibinfo{title}{Microlenses arrays: {{Fabrication}}, materials, and applications}.
\newblock \emph{\bibinfo{journal}{Microscopy Research and Technique}} \textbf{\bibinfo{volume}{84}}, \bibinfo{pages}{2784--2806} (\bibinfo{year}{2021}).

\bibitem{beguelin_commented_2021}
\bibinfo{author}{B{\'e}guelin, J.}, \bibinfo{author}{Voelkel, R.} \& \bibinfo{author}{Scharf, T.}
\newblock \bibinfo{title}{Commented review on refractive microlenses and microlens arrays metrology}.
\newblock \emph{\bibinfo{journal}{Journal of Optical Microsystems}} \textbf{\bibinfo{volume}{1}}, \bibinfo{pages}{030901} (\bibinfo{year}{2021}).

\bibitem{li_focal_1981}
\bibinfo{author}{Li, Y.} \& \bibinfo{author}{Wolf, E.}
\newblock \bibinfo{title}{Focal shifts in diffracted converging spherical waves}.
\newblock \emph{\bibinfo{journal}{Optics Communications}} \textbf{\bibinfo{volume}{39}}, \bibinfo{pages}{211--215} (\bibinfo{year}{1981}).

\bibitem{bouwkamp_diffraction_1954}
\bibinfo{author}{Bouwkamp, C.~J.}
\newblock \bibinfo{title}{Diffraction {{Theory}}}.
\newblock \emph{\bibinfo{journal}{Reports on Progress in Physics}} \textbf{\bibinfo{volume}{17}}, \bibinfo{pages}{35--100} (\bibinfo{year}{1954}).

\bibitem{stratton_diffraction_1939}
\bibinfo{author}{Stratton, J.~A.} \& \bibinfo{author}{Chu, L.~J.}
\newblock \bibinfo{title}{Diffraction {{Theory}} of {{Electromagnetic Waves}}}.
\newblock \emph{\bibinfo{journal}{Physical Review}} \textbf{\bibinfo{volume}{56}}, \bibinfo{pages}{99--107} (\bibinfo{year}{1939}).

\bibitem{born_principle_1999}
\bibinfo{author}{Born, M.} \& \bibinfo{author}{Wolf, E.}
\newblock \emph{\bibinfo{title}{Principle of {{Optics}}}} \bibinfo{edition}{7th} edn (\bibinfo{publisher}{Cambridge}, \bibinfo{year}{1999}).

\bibitem{kottler_elektromagnetische_1923}
\bibinfo{author}{Kottler, F.}
\newblock \bibinfo{title}{Elektromagnetische {{Theorie}} der {{Beugung}} an schwarzen {{Schirmen}}}.
\newblock \emph{\bibinfo{journal}{Annalen der Physik}} \textbf{\bibinfo{volume}{376}}, \bibinfo{pages}{457--508} (\bibinfo{year}{1923}).

\bibitem{aieta_out--plane_2012}
\bibinfo{author}{Aieta, F.} \emph{et~al.}
\newblock \bibinfo{title}{Out-of-{{Plane Reflection}} and {{Refraction}} of {{Light}} by {{Anisotropic Optical Antenna Metasurfaces}} with {{Phase Discontinuities}}}.
\newblock \emph{\bibinfo{journal}{Nano Letters}} \textbf{\bibinfo{volume}{12}}, \bibinfo{pages}{1702--1706} (\bibinfo{year}{2012}).

\bibitem{li_dependence_1982}
\bibinfo{author}{Li, Y.}
\newblock \bibinfo{title}{Dependence of the focal shift on {{Fresnel}} number and f number}.
\newblock \emph{\bibinfo{journal}{JOSA}} \textbf{\bibinfo{volume}{72}}, \bibinfo{pages}{770--774} (\bibinfo{year}{1982}).

\bibitem{wang_far-zone_1995}
\bibinfo{author}{Wang, W.} \& \bibinfo{author}{Wolf, E.}
\newblock \bibinfo{title}{Far-zone behavior of focused fields in systems with different {{Fresnel}} numbers}.
\newblock \emph{\bibinfo{journal}{Optics Communications}} \textbf{\bibinfo{volume}{119}}, \bibinfo{pages}{453--459} (\bibinfo{year}{1995}).

\bibitem{aieta_aberrations_2013}
\bibinfo{author}{Aieta, F.}, \bibinfo{author}{Genevet, P.}, \bibinfo{author}{Kats, M.} \& \bibinfo{author}{Capasso, F.}
\newblock \bibinfo{title}{Aberrations of flat lenses and aplanatic metasurfaces}.
\newblock \emph{\bibinfo{journal}{Optics Express}} \textbf{\bibinfo{volume}{21}}, \bibinfo{pages}{31530--31539} (\bibinfo{year}{2013}).

\bibitem{elsawy_multiobjective_2021}
\bibinfo{author}{Elsawy, M. M.~R.} \emph{et~al.}
\newblock \bibinfo{title}{Multiobjective {{Statistical Learning Optimization}} of {{RGB Metalens}}}.
\newblock \emph{\bibinfo{journal}{ACS Photonics}} \textbf{\bibinfo{volume}{8}}, \bibinfo{pages}{2498--2508} (\bibinfo{year}{2021}).

\end{thebibliography}


\begin{thebibliography}{1}
\expandafter\ifx\csname url\endcsname\relax
  \def\url#1{\burl{#1}}\fi
\expandafter\ifx\csname urlprefix\endcsname\relax\def\urlprefix{URL }\fi
\providecommand{\bibinfo}[2]{#2}
\providecommand{\eprint}[2][]{\url{#2}}
\providecommand{\doi}[1]{\url{https://doi.org/#1}}
\bibcommenthead

\bibitem{li_focal_2005}
\bibinfo{author}{Li, Y.}
\newblock \bibinfo{title}{Focal shifts in diffracted converging electromagnetic waves. {{I}}. {{Kirchhoff}} theory}.
\newblock \emph{\bibinfo{journal}{Journal of the Optical Society of America A}} \textbf{\bibinfo{volume}{22}}, \bibinfo{pages}{68--76} (\bibinfo{year}{2005}).

\bibitem{richards_electromagnetic_1959}
\bibinfo{author}{Richards, B.} \& \bibinfo{author}{Wolf, E.}
\newblock \bibinfo{title}{Electromagnetic diffraction in optical systems, {{II}}. {{Structure}} of the image field in an aplanatic system}.
\newblock \emph{\bibinfo{journal}{Proceedings of the Royal Society of London. Series A. Mathematical and Physical Sciences}} \textbf{\bibinfo{volume}{253}}, \bibinfo{pages}{358--379} (\bibinfo{year}{1959}).

\bibitem{stratton_diffraction_1939}
\bibinfo{author}{Stratton, J.~A.} \& \bibinfo{author}{Chu, L.~J.}
\newblock \bibinfo{title}{Diffraction {{Theory}} of {{Electromagnetic Waves}}}.
\newblock \emph{\bibinfo{journal}{Physical Review}} \textbf{\bibinfo{volume}{56}}, \bibinfo{pages}{99--107} (\bibinfo{year}{1939}).

\end{thebibliography}

\end{document}


\title[Supplementary Materials: Diffraction-limited operation of micro-metalenses]{Supplementary Materials: Diffraction-limited operation of micro-metalenses: fundamental bounds and designed rules for pixel integration}

\author[1]{\fnm{Nicolas}  \sur{Kossowski} \email{nicolas.kossowski@crhea.cnrs.fr}}
\author[1]{ \fnm{Christina} \sur{Kyrou}}
\author[1]{ \fnm{Rémi} \sur{Colom}}
\author[1]{\fnm{Pierre-Marie} \sur{Coulon}}
\author[1]{\fnm{Virginie} \sur{Br\"{a}ndli}}
\author[2]{\fnm{Jean-Luc} \sur{Reverchon}}
\author[1]{\fnm{Samira} \sur{Khadir}}
\author[1,3]{\fnm{Patrice} \sur{Genevet} \email{patrice.genevet@mines.edu}}

\affil[1]{\orgdiv{Université Côte d'Azur}, \orgname{CNRS, CRHEA}, \orgaddress{\street{Rue Bernard Gregory}, \city{Valbonne}, \postcode{06560}, \country{France}}}
\affil[2]{\orgdiv{III-V Lab}, \orgaddress{\street{1 Avenue Augustin Fresnel}, \city{Palaiseau}, \postcode{91767}, \country{France}}}
\affil[3]{\orgdiv{Physics department}, \orgname{Colorado School of Mines}, \orgaddress{\street{1523 Illinois Street}, \city{Golden, CO}, \postcode{80401}, \country{USA}}}

\maketitle

\section{Vectorial Diffraction by a Metalens}\label{sec:vectorial_diff}
\begin{annotateblock}
    In this appendix, we solve the vectorial form of the light diffraction through an apertured metalens. The metalens diameter, with radius $a$, defines the aperture size. The metalens is illuminated by a normally incident, $\bm{k}_{i,\perp}=0$, and $x$-linearly polarized plane wave. Applying the general law of refraction, Eq.~\eqref{main-eq:generalized_law_refraction} of the main text, in the plane of the metasurface, the transverse wavevector $\bm{k}_{a,\perp}$ at the output of the metasurface is exactly the gradient of the phase profile $\varphi(\bm{r}^\prime)$,
    \begin{equation}
	\bm{k}_{a,\perp} = \bm{D}\varphi,
	\label{eq:generalized_law_refraction_incidence}
    \end{equation}
    where $\bm{D} = [\partial\varphi/\partial x \; \partial\varphi/\partial y]$ and in the particular case of a metalens, it is a conservative vector field, thus it is the gradient $\bm{D}\varphi = \nabla \varphi$. For a metalens, where $\varphi(\bm{r}^\prime)$ is given by Eq.~\eqref{main-eq:ph_lens} of the main text, it yields
    \begin{equation}
	k_x = - \frac{2\pi}{\lambda} \frac{x^\prime}{\sqrt{x^{\prime 2} + y^{\prime 2} + f^2}}
	\quad\text{and}\quad
	k_y = - \frac{2\pi}{\lambda} \frac{y^\prime}{\sqrt{x^{\prime 2} + y^{\prime 2} + f^2}}.
	\label{eq:kxky}
    \end{equation}
    Defining the azimuthal angle $\alpha$ by
    \begin{equation}
	\cos\alpha = \frac{x^\prime}{\sqrt{x^{\prime 2} + y^{\prime 2}}}
	\quad\text{and}\quad
	\sin\alpha = \frac{y^\prime}{\sqrt{x^{\prime 2} + y^{\prime 2}}},
	\label{eq:cosa_sina}
    \end{equation}
    as well as the angle $\beta$ by
    \begin{equation}
	\cos\beta = \frac{f}{\sqrt{x^{\prime 2} + y^{\prime 2} + f^2}},
	\quad\text{and}\quad
	\sin\beta = \frac{\sqrt{x^{\prime 2} + y^{\prime 2}}}{\sqrt{x^{\prime 2} + y^{\prime 2} + f^2}}.
	\label{eq:cosb_sinb}
    \end{equation}
    we can rewrite Eq.~\eqref{eq:kxky} in a simple form,
    \begin{equation}
	\bm{k} = k
	\begin{bmatrix}
	    - \cos\alpha\sin\beta\\
	    - \sin\alpha\sin\beta\\
	       \cos\beta
	\end{bmatrix},
	\label{eq:kxkykz}
    \end{equation}
    where we used $k_z = \sqrt{k^2 - k_{a,\perp}^2}$. We remark that for $\sqrt{x^{\prime 2} + y^{\prime 2}} \leq a$, then $\sin\beta \leq \text{NA}$. By analogy to the geometrical method developed in ref.~\cite{li_focal_2005}, we can express the electric field $\bm{E}_a$ inside the aperture as follow,
    \begin{equation}
	\bm{E}_a(\bm{r}^\prime) = t(\bm{r}^\prime) \annotatebis{E_a}
	\begin{bmatrix}
	    \cos^2\alpha\cos\beta + \sin^2\alpha\\
	    -\cos\alpha\sin\alpha(1-\cos\beta)\\
	    \cos\alpha\sin\beta
	\end{bmatrix}
	e^{i\varphi(\bm{r}^\prime)},
	\label{eq:Ea_vector}
    \end{equation}
    were $t(\bm{r}^\prime)$ is the transmission. We can verify that the electric field is transverse by satisfying the local condition $\bm{E}_a\cdot\bm{k} = 0$ at every point of the aperture plane. In consequence, the electric field is no longer polarized along $x$, but it has non-zero $y$ and $z$ components.  When the metalens has a small numerical aperture, the $y$ and $z$ components of the electric $\bm{E}_a$ field become negligible. From Eq.~\eqref{eq:Ea_vector}, if $\text{NA} < 0.4$ then $\cos \beta < 0.1$ and so the $E_{a,y}$ component can be neglected in front of the $E_{a,x}$ and $E_{a,z}$. For $\text{NA} < 0.1$ both $E_{a,y}$ and $E_{a,z}$ components can be neglected in front of the $E_{a,x}$.
    
    The magnetic field $\bm{H}_a(\bm{r}^\prime)$ can be derived as 
    \begin{equation}
	\bm{H}_a(\bm{r}^\prime) = \unitv{k}\times\bm{E}_a(\bm{r}^\prime)
	\quad\text{and}\quad
	\unitv{k} = \frac{\bm{k}}{k}.
	\label{eq:Ha_vector}
    \end{equation}
    \begin{annotateblockbis}
         Considering power of incident on the aperture by the incident plane wave $(\bm{E}_i, \bm{H}_i)$, it is given by the flux of the time-averaged Poyting vector $\bm{S}_i$ through the surface of the metasurface, i.e,
         \begin{equation}
             P_0 = \iint_\text{ML} \bm{S}_i\cdot\unitv{z} dS = \frac{c}{8} a^2 \abs{E_0}^2
             \quad\text{with}\quad
             \bm{S}_i = \frac{c}{8\pi} \real[\bm{E}_i\times\bm{H}_i^*]
         \end{equation}
         Just after the metasurface, using Eq.~\eqref{eq:kxkykz}-\eqref{eq:Ha_vector}, the $z$-component of the time-averaged Poynting vector $\bm{S}_a$, can be found to be
         \begin{equation}
             \bm{S}_a\cdot\unitv{z} = \frac{c}{8\pi} \abs{E_a}^2 \cos\beta,
         \end{equation}
         Consequently, in order to conserve the energy through the metalens, i.e.
         \begin{equation}
             \iint_\text{ML} \bm{S}_a\cdot\unitv{z} dS = P_0,
         \end{equation}
         the amplitude of the electric field $\bm{E}_a$ has to be renormalized by a factor $1/\cos\beta$ \cite{li_focal_2005, richards_electromagnetic_1959},
         \begin{equation}
             E_a = \frac{E_0}{\sqrt{\cos\beta}}.
         \end{equation}
    \end{annotateblockbis}
    By replacing $(\bm{E}_a, \bm{H}_a)$ by their expressions Eq.~\eqref{eq:Ea_vector} and Eq.~\eqref{eq:Ha_vector}, respectively, in the equivalence theorem, Eq.~\eqref{main-eq:equivTh} of the main text, it is possible to compute the diffracted electric field $\bm{E}_d$ by the metalens.

    We emphasize that metasurfaces are generally surrounded by a transparent medium or disposed on a substrate, which contributes to the field distribution in the vicinity of the metasurface. The convergence of the integrals in \annotate{Eq.~}\eqref{main-eq:equivTh} can be enforced by considering that the incident waves are bounded with a finite spatial distribution, such as a Gaussian beam.

    On the optical axis, the $y$ and $z$ components of the electric fields cancel out by symmetry. For example, the component $E_{d,z}$ is positive in the half-space $x>0$, since $\cos\alpha > 0$, and negative otherwise. Therefore, both contributions cancel out on the optical axis. Similar conclusions can be drawn for $E_{d,y}$. \annotatebis{Moreover, for numerical aperture $\text{NA}$ smaller than $0.7$, then the normalization factor can be approximated by $1$, i.e. $\sqrt{\cos\beta} > 0.8$.} In this case, we can consider that the field in the aperture is equal to the incident plane wave phase-shifted by the metasurface. Equation~\eqref{main-eq:equivTh} of the main text becomes the Rayleigh-Sommerfeld integrals evaluated on the optical axis,
    \begin{equation}
	\label{eq:Rayleigh-Sommerfeld}
	   E_{d,x}(\bm{r}) \annotatebis{\approx} -\frac{E_0}{4\pi}\iint_A t(\bm{r}^\prime) \left[ \left( ik - \frac{1}{s}\right) \frac{z}{s} + ik \right]\frac{e^{iks}}{s} e^{i\varphi(\bm{r}^\prime)}d\bm{r}^\prime
    \end{equation}
    with $s = \sqrt{\rho^{ 2} + z^2}$ and $\rho = \sqrt{x^{\prime 2} + y^{\prime 2}}$.

    Equation~\eqref{eq:Rayleigh-Sommerfeld}, indicates that the diffracted field on the optical axis, is the convolution product of the complex transmission $t(\bm{r}^\prime)e^{i\varphi(\bm{r}^\prime)}$ with the field radiated by the fictious current densities inside the aperture. As a consequence, the wavevectors are spread by the aperture function.
\end{annotateblock}

\section{Diffraction by a Plain Aperture}\label{sec:ap_circAperture}
We start from \annotate{Eq.~}\eqref{eq:Rayleigh-Sommerfeld}, in the case where $\varphi(\bm{r}^\prime) = 0$ \annotate{and $t(\bm{r}^\prime) = 1$} for $\bm{r}^\prime\in A$. Considering a point $\bm{r}$ located on the optical axis $z$, the $x$-component of the electric field is given by
\begin{equation}
    \begin{aligned}
    E_{d,x}(z) = \annotate{-}\frac{E_0}{4\pi}\iint_A & \left[ \left( ik-\frac{1}{\sqrt{\rho^2+z^2}} \right)\frac{z}{\sqrt{\rho^2+z^2}} + ik\right]\\
    &\times\frac{e^{ik\sqrt{\rho^2+z^2}}}{\sqrt{\rho^2+z^2}} \rho d\rho d\theta.
    \end{aligned}
    \label{eq:ap_Ex}
\end{equation}
The surface integral reduces to a single integral by invariance of rotation around the $z$. If the following, we adopt the \annotate{tilde} notation for normalized distance by the radius of the aperture, $\tilde{z} = z/a$. Thus we obtain
\begin{equation}
    \begin{aligned}
    E_{d,x}(\tilde{z}) = \annotate{-}\frac{E_0}{\annotate{2}}\int_0^1&\left[ \left( iak-\frac{1}{\sqrt{\tilde{\rho}^2+\tilde{z}^2}} \right)\frac{\tilde{z}}{\sqrt{\tilde{\rho}^2+\tilde{z}^2}} + iak\right]\\
    \label{eq:ap_Ex_norm}
    &\times\frac{e^{iak\sqrt{\tilde{\rho}^2+\tilde{z}^2}}}{\sqrt{\tilde{\rho}^2+\tilde{z}^2}} \tilde{\rho} d\tilde{\rho}.
    \end{aligned}
\end{equation}
Now, we are looking for a closed formula of \annotate{Eq.~}\eqref{eq:ap_Ex_norm}. We observe that in the integrand, the term in the bracket is the summation of two terms. The first term is proportional to $1/\tilde{z}$, and the second term is a constant. Consequently, for $\tilde{z}$ sufficiently large, the term in the bracket is considered constant and can be taken out of the integral. We plotted the amplitude function of the term in the bracket and found that approximately for $\tilde{z}>1$ we have a good approximation. Therefore, for $\tilde{z}>1$,
\begin{equation}
   \label{eq:ap_Ex_approx}
   E_{d,x}(\tilde{z}) \approx \annotate{-}E_0\annotate{iak}\int_0^1 \frac{e^{iak\sqrt{\tilde{\rho}^2+\tilde{z}^2}}}{\sqrt{\tilde{\rho}^2+\tilde{z}^2}} \tilde{\rho} d\tilde{\rho},
\end{equation}
which can be easily integrated by observing the integrand in the form $u^\prime \exp u$. The intensity is then obtained in the form
\begin{eqnarray}
    \label{eq:ap_circAp_cos}
    I = 2 I_0\left\{ 1 - \cos\left[ ak\left(\sqrt{1+\tilde{z}^2}- \tilde{z}\right) \right]\right\},
\end{eqnarray}
with $I_0 = \abs{E_0}^2$. Using the trigonometric identity $2\sin^2\theta = 1-\cos2\theta$, \annotate{Eq.~}\eqref{eq:ap_circAp_cos} is transformed into
\begin{eqnarray}
    \label{eq:ap_circAp_sin}
    I = 4 I_0 \sin^2\left[ \frac{ak}{2}\left(\sqrt{1+\tilde{z}^2}- \tilde{z}\right) \right],
\end{eqnarray}
which can be approximated into \annotate{Eq.~}\eqref{main-eq:int_circAperture} using a Taylor approximation square root function. \annotate{To find the focal} spot of the aperture, \annotate{we} consider the \annotate{derivative} of the intensity~\eqref{eq:ap_circAp_cos},
\begin{eqnarray}
    \label{eq:ap_der_I}
    \frac{dI}{d\tilde{z}} = 2 I_0 \frac{\tilde{z}-\sqrt{1+\tilde{z}^2}}{\sqrt{1+\tilde{z}^2}} \sin\left[ ak\left(\sqrt{1+\tilde{z}^2}- \tilde{z}\right) \right]
\end{eqnarray}
The maxima and minima, are obtained for $dI/d\tilde{z} = 0$, and so when $\sin\left[ ak\left(\sqrt{1+\tilde{z}^2}- \tilde{z}\right) \right]= 0$, which means that
\begin{equation}
    \tilde{z}_p = \frac{ak}{2p\pi} - \frac{p\pi}{2ak} \;\text{for}\; p\in \mathbb{N}^*.
\end{equation}
The intensity is maximal for $p=1$ and $\tilde{z}_1 > \tilde{z}_{p\neq 1}$.

\section{Diffraction by a Lens}\label{sec:circLens}
By analogy to \ref{sec:ap_circAperture}, we derive the \annotate{approximated} $x$-component of the electric field on the optical axis for $\tilde{z}>1$ \annotate{using Eq.~\eqref{eq:Rayleigh-Sommerfeld}}
\begin{equation}
    E_{d,x}(z) = \annotate{-}\annotate{E_0} \int_0^1 iak \frac{e^{iak\sqrt{\tilde{\rho}^2+\tilde{z}^2}}}{\sqrt{\tilde{\rho}^2+\tilde{z}^2}}e^{iak(f-\sqrt{\tilde{\rho}^2+f^2})}  \tilde{\rho}d\tilde{\rho}.
\end{equation}
We consider also that $\tilde{f}>1$. We approximate the term $1/\sqrt{\tilde{\rho}^2+\tilde{z}^2}$ by $1/\tilde{z}$ and we expand the term in the exponential by a Taylor series,
\begin{equation}
    E_{d,x}(z) = \annotate{-}E_0 \annotate{iak} \frac{e^{iak\tilde{z}}}{\tilde{z}} \int_0^1 e^{i\frac{ak}{2}\tilde{\rho}^2(\frac{1}{\tilde{z}}-\frac{1}{\tilde{f}})}  \tilde{\rho}d\tilde{\rho}.
\end{equation}
We recognize the integrand is in the form $\annotate{2} u^\prime u \exp u^2$, so we can integrate and obtain
\begin{eqnarray}
    E_{d,x}(z) = \annotate{-}E_0 \frac{iak}{2} \frac{e^{iak\tilde{z}}}{\tilde{z}} \sinc\left[\frac{ak}{4}\left(\frac{1}{\tilde{z}}-\frac{1}{\tilde{f}}\right)\right]\annotate{e^{-i\frac{ak}{4}(\frac{1}{\tilde{z}}-\frac{1}{\tilde{f}})}},
\end{eqnarray}
which leads to \annotate{Eq.~}\eqref{main-eq:int_circLens} of the main text.
Using the definition of the Fresnel number~\eqref{main-eq:fresnelNb} in the manuscript, we can rewrite naturally the intensity as
\begin{equation}
    I(\tilde{z}) = I_0 \frac{a^2k^2}{4\tilde{z}^2} \sinc^2\left[ \frac{ak}{4\tilde{z}}-\frac{\pi N}{2} \right],
\end{equation}
where it appears that for low Fresnel number, the intensity tends to the solution of the problem of the plain aperture.

\section{Focusing Efficiency}\label{sec:focal_efficiency}
\begin{annotateblock}

We define the focusing efficiency as the ratio between the optical power passing through a square pixel with size $\SI{10}{\micro\metre}\times\SI{10}{\micro\metre}$ divided by the optical power transmitted through the metasurface,
\begin{annotateblockbis}
    \begin{equation}
    \eta = \frac{\iint_{\text{pixel}} \bm{S}(\bm{r})\cdot\unitv{z} ds}{P_0},
    \label{eq:focalization_efficiency}
\end{equation}
\end{annotateblockbis}
where $\bm{S}$ is the \annotatebis{time-averaged  Poynting vector at $\bm{r}$}. The focusing efficiency quantifies the spreading (and absorption, if any) of light after transmission through the metalens, thus estimating both the design wavefront accuracy and the optical crosstalk between pixels. We computed the focusing efficiency using the vectorial diffraction theory, where the magnetic field is given by \cite{stratton_diffraction_1939}
\begin{equation}
    \begin{aligned}
        \bm{H}_d(\bm{r}) = \frac{1}{4\pi} \iint_A \bigg\{ & ik\left[\unitv{n}\times\bm{E}_a(\bm{r}^\prime)\right]G(\bm{r},\bm{r}^\prime) \\
        & - \left[\unitv{n}\times\bm{H}_a(\bm{r}^\prime)\right]\times\nabla G(\bm{r},\bm{r}^\prime) \\
        & - \left[\unitv{n}\cdot\bm{H}_a(\bm{r}^\prime)\right]\nabla G(\bm{r},\bm{r}^\prime)\bigg\}d \bm{r}^\prime\\
        & + \frac{1}{4\pi k} \oint_C \nabla G(\bm{r},\bm{r}^\prime) [\bm{E}_a(\bm{r}^\prime)\cdot \unitv{l}] d \bm{r}^\prime,
    \end{aligned}
    \label{eq:equivTh_H}
\end{equation}

\begin{figure*}[!ht]
    \centering
    \includegraphics{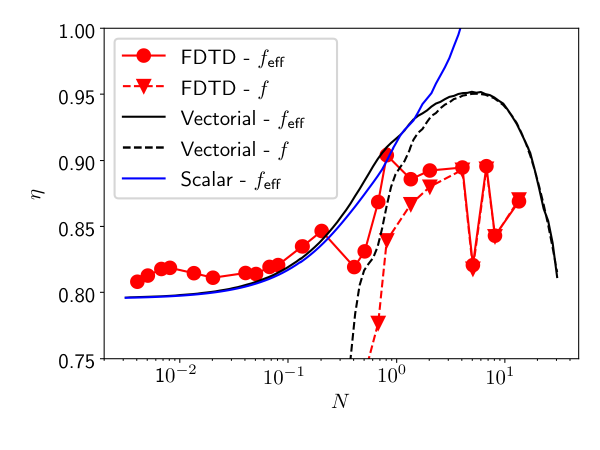}
    \caption{\annotate{Focusing efficiency $\eta$ as a function of the Fresnel number $N$ computed via vectorial diffraction theory (black color), \annotatebis{scalar diffraction theory (blue)} and FDTD simulations (red color). The focusing efficiencies have been computed in two different planes, at the plane of the maximum of intensity, i.e. effective focal distance $f_{\text{eff}}$ represented by the continuous lines, and at the the plane associated with the designed focal distance $f$. For both theory and FDTD, the focusing efficiency drops considerably for Fresnel numbers smaller than 4 at $z=f$. This is a consequence of the focal shift and the large divergence of the beam after the focal spot. For smaller Fresnel numbers, the focusing efficiency converges toward a finite value of approximately $0.79$.}}
    \label{fig:fig5}
\end{figure*}

\begin{annotateblockbis}
    We compare the vectorial theory with a scalar diffraction theory for which we consider inside the aperture the electric field $\bm{E}_{a,\text{sc.}}$ is simply given the incident field phase shifted by the metasurface,
    \begin{equation}
        \bm{E}_{a,\text{sc.}} = \bm{E}_i e^{i\varphi(\bm{r})}, \quad\text{and}\quad \bm{H}_{a,\text{sc.}} = \bm{H}_i e^{i\varphi(\bm{r})}.
    \end{equation}
    Therefore, the energy  is conserved while transmitted by the metasurface. As plotted in Fig.~\ref{fig:fig5}, we can observe that for small Fresnel number, the scalar theory agrees well with the vectorial theory but for large Fresnel number, i.e. large numerical aperture when the diameter is fixed, the theory breaks down with a focusing efficiency larger than 1. It results from non-taking into account the contribution of the other component of the electromagnetic field.
\end{annotateblockbis}

The FDTD simulations were performed using the commercial software ANSYS Lumerical. We simulated the whole metalens on top of the sapphire substrate. To define the apertures, we decided to cover the part of the substrate outside of the metasurface area with a perfect electrical conductor (PEC). This avoids diffraction by the bohundaries of the simulation domains, window effect, and enables simple normalization of the input power on the metasurface aperture, simply because the rest of the light is reflecting at the PEC region. The entire simulation domain was bounded by perfectly matched layers (PML). The electric field at the focal spot was computed using the exact far-field projection method of light measured at a monitor placed directly at the metasurface transmission plane. The magnetic field was computed using Maxwell's equation $\nabla\times\bm{E} = i\mu_0\omega \bm{H}$, where the derivatives of $\bm{E}$ were computed from two consecutive but spatially shifted far-field projections.

From Fig.~\ref{fig:fig5}, on the one hand, if the pixel is placed in the real focal plane, the overall focusing efficiency of the metasurface is comparable to the theoretical values, as it follows the same trend. As we could expect, however, the focusing efficiency obtained by FDTD simulations is smaller. We attribute this difference to the sampling of the phase profile and potential coupling between the elements. We notice that even for large Fresnel numbers, the focusing efficiency may drop, but it may be improved considerably by optimizing both the position and the size of the elements. On the other hand, neglecting the focal shift and placing the detector in the wrong plane, the focusing efficiency converges to zero. Consequently, the gain in performance after adding a focusing metasurface relies on precise manufacturing and assembly to control the distance and the positioning of the pixel with respect to the micro-metalens optical axis and effective focal distance.
\end{annotateblock}

\section{Sampling Condition}\label{sec:sampling}
We established a sampling condition on the phase profile of the metasurface that requires at least $N$ meta-atoms per Fresnel zone. A Fresnel zone is the spatial region defined by the wrapped phase profile varying continuously from $0$ to $2\pi$. To contain at least $N_m$ meta-atoms per Fresnel zone, the phase difference between two consecutive meta-atoms requires
\begin{equation}
    \Delta \varphi < \frac{2\pi}{N_m}.
\end{equation}
In the case of a metalens, we consider the phase difference between two elements located at the border of the metalens. To have proper sampling of the Fresnel zone at this location therefore,
\begin{equation}
    \abs{\varphi(a) - \varphi(a-p)} < \frac{2\pi}{N_m}.
    \label{eq:phase_diff}
\end{equation}
Using \annotate{Eq.~}\eqref{main-eq:ph_lens} of the main text, we obtain
\begin{equation}
    k \abs{ \sqrt{a^2 + f^2} - \sqrt{ (a-p)^2 + f^2} } < \frac{2\pi}{N_m}.
\end{equation}
 Assuming we can neglect the term $p^2$ in front of other terms, we can approximate the second square roots by
 \begin{equation}
     \sqrt{ (a-p)^2 + f^2} \approx \sqrt{a^2 + f^2} (1-\frac{ap}{a^2 + f^2}),
 \end{equation}
 leading to
 \begin{equation}
     \abs{\varphi(a) - \varphi(a-p)} \approx \frac{2\pi p}{\lambda} \frac{a}{\sqrt{a^2 + f^2}}.
 \end{equation}
 We recognize that $\text{NA} = a / \sqrt{a^2 + f^2}$, and therefore we can obtain \annotate{Eq.~}\eqref{main-eq:na_p} in the main text.

 \bibliography{ref}